\documentclass[12pt,a4paper]{article}

\usepackage[nosort]{cite}
\usepackage[hyperref,bulletsep]{collect}
\usepackage{amsmath,amssymb,graphicx}
\usepackage{pst-all}
\usepackage{bbm}

\makeatletter \@addtoreset{equation}{section} \makeatother

\makeatletter
\let\old@startsection=\@startsection
\let\oldl@section=\l@section
\renewcommand{\@startsection}[6]{\old@startsection{#1}{#2}{#3}{#4}{#5}{#6\mathversion{bold}}}
\renewcommand{\l@section}[2]{\oldl@section{\mathversion{bold}#1}{#2}}
\makeatother

\makeatletter
\let\old@makecaption=\@makecaption
\def\@makecaption{\small\old@makecaption}
\makeatother

\renewcommand{\geq}{\geqslant}

\begin{document}

\thispagestyle{empty}
\begin{flushright}\footnotesize
\texttt{LPTENS-08/06}\\
 \texttt{UUITP-01/08} \vspace{0.8cm}
\end{flushright}

\renewcommand{\thefootnote}{\fnsymbol{footnote}}
\setcounter{footnote}{0}

\begin{center}
{\Large\textbf{\mathversion{bold} Quantum Giant Magnons}\par}

\vspace{1.5cm}

\textrm{K.~Zarembo\footnote{Also at ITEP, Moscow, Russia}}
\vspace{8mm}

\textit{Department of Physics and Astronomy, Uppsala University\\
SE-751 08 Uppsala, Sweden}\\
\texttt{Konstantin.Zarembo@teorfys.uu.se} \vspace{3mm}

\textit{Laboratoire de Physique Th\'eorique de l'Ecole Normale
Sup\'erieure\\ 24 rue Lhomond, Paris CEDEX 75231, France }\\
\vspace{3mm}


\par\vspace{1cm}

\textbf{Abstract} \vspace{5mm}

\begin{minipage}{14cm}
The giant magnons are classical solitons of the $O(N)$ sigma-model,
which play an important role in the AdS/CFT correspondence. We study
quantum giant magnons first at large $N$ and then exactly using
Bethe Ansatz, where giant magnons can be interpreted as holes in the
Fermi sea. We also identify a solvable limit of Bethe Ansatz in
which it describes a weakly-interacting Bose gas at zero
temperature. The examples include the $O(N)$ model at large-$N$,
weakly interacting non-linear Schr\"odinger model, and nearly
isotropic XXZ spin chain in the magnetic field.
\end{minipage}

\end{center}

\vspace{0.5cm}


\newpage
\setcounter{page}{1}
\renewcommand{\thefootnote}{\arabic{footnote}}
\setcounter{footnote}{0}

\section{Introduction}

The giant magnons \cite{Hofman:2006xt} are solitons on the string
world-sheet in $AdS_5\times S^5$ and are argued to be the
fundamental building blocks of the spectrum in the AdS/CFT
correspondence. One of the remarkable features of giant magnons is
periodicity of their momentum, which has a geometric origin
\cite{Hofman:2006xt}. This periodicity is quite puzzling since the
centre of mass of a giant magnon, the collective coordinate
canonically conjugate to the momentum, should then be quantized,
pointing perhaps to some underlying lattice structure in the sigma
model on $AdS_5\times S^5$. The semiclassical quantization of the
giant magnon was carried out in
\cite{Minahan:2007gf,Papathanasiou:2007gd,Chen:2007vs,Gromov:2008ie,Heller:2008at}.
The purpose of this paper is to go beyond the semiclassical
approximation, albeit not in string theory in $AdS_5\times S^5$. The
prime example will be the $O(N)$ sigma-model, which also admits
giant magnons as classical solutions. Following
\cite{Tyupkin:1975pk,Kulish:1976ek}, we will identify quantum giant magnons in the
$O(N)$ model with the holes in the Fermi sea of the fundamental
vector particles. The Fermi sea arises in the exact Bethe-Ansatz
solution of the model
\cite{Zamolodchikov:1978xm,Polyakov:1983tt,Hasenfratz:1990zz,Hasenfratz:1990ab}.

The analogy with \cite{Tyupkin:1975pk,Kulish:1976ek}  is possible because the
solitons of the non-linear Schr\"odinger (NLS) equation considered
there have much in common with giant magnons. Both are particular
examples of {\it dark solitons} \cite{darks}. A dark soliton can be
pictured as a dark spot moving through a bright medium (hence the
name) or, more appropriately in the present context, as a localized
dilution of the Bose-Einstein condensate. It is characterized by two
conditions: (i) finite background density: $\phi \rightarrow \phi
\,{\rm e}\,^{-i\mu t}$, $\left\langle \phi \right\rangle\neq 0$
($\phi $ is the field that carries the condensed charge, $\mu $ is
the chemical potential)\footnote{More precisely, this condition
states that the field has the form $\left\langle \phi
\right\rangle\,{\rm e}\,^{-i\mu t}$ asymptotically at spacial
infinity.}; and (ii) twisted boundary conditions: $\phi (+\infty
,t)=\,{\rm e}\,^{i\Delta \varphi }\phi (-\infty ,t)$. In other
words, the phase of $\phi $ experiences a finite increment as one
crosses the soliton, and the modulus of $\phi $ has a dip in the
soliton's core.

We will consider giant magnon solutions (which belong to the class
of dark solitons described above) in the $O(N)$ sigma model:
\begin{equation}\label{on}
 S=\frac{N}{2\lambda }\int_{}^{}d^2x\,\partial ^\nu \mathbf{n}\cdot \partial
 _\nu \mathbf{n},
\end{equation}
where $\mathbf{n}$ is an $N$-dimensional unit vector. The giant
magnon is a soliton on $S^2$, which in terms of the charged field
$\phi =\sin\vartheta \,{\rm e}\,^{i\varphi }$, where
$\mathbf{n}=(\sin\vartheta \cos\varphi, \sin\vartheta\sin\varphi
,\cos\vartheta,\mathbf{0} )$, has the form \cite{Hofman:2006xt}
\begin{equation}\label{HM}
 \phi =\,{\rm e}\,^{-i\mu t}\left(v-i\sqrt{1-v^2}
 \tanh\frac{\mu (x-vt)}{\sqrt{1-v^2}}\right).
\end{equation}
The solution obviously satisfies the above conditions (i) and (ii).

The giant magnon solution is strikingly similar to the dark soliton
\cite{tsuzuki,zakharov-shabat}
\begin{equation}\label{dark}
 \phi =\frac{1}{2\sqrt{g}}\left[
 v-i\sqrt{2\mu -v^2}\tanh\frac{ \sqrt{2\mu -v^2}(x-vt)}{2}
 \right]
 \end{equation}
of the NLS model\footnote{The time-dependent phase in $\phi $ here
is traded for the chemical potential in the Lagrangian. }:
\begin{equation}\label{NLS}
 S_{\rm NLS}=\int_{}^{}dt\,dx\,\left(
 i\phi ^*\dot{\phi }-|\acute{\phi }|^2-g|\phi |^4+\mu |\phi
 |^2\right).
\end{equation}
An apparent difference between the the giant magnon and the NLS
soliton is that the size of the latter can be arbitrary large and
actually becomes infinite at $v^2=2\mu $, while the size of the
giant magnon depends on the velocity only through the trivial
Lorentz-contraction factor and is always smaller than $1/\mu $. We
will see in sec.~\ref{largeN} that this is an artifact of the
classical approximation. The quantum giant magnon also has a
variable, velocity-dependent size which turns to infinity when
soliton moves at the speed of sound.

 In the
framework of Bethe Ansatz the ground state of the quantum NLS model
at non-zero chemical potential is represented by a Fermi sea of
interacting particles \cite{Lieb:1963rt}, 1d bosons with a local
repulsive interaction for which (\ref{NLS}) is the second-quantized
action. Because of the repulsion particles in some sense obey the
Fermi statistics. The spectrum of elementary excitations has two
branches, the particles and the holes. At weak coupling ($g\ll
\sqrt{\mu }$) the spectral properties of the hole excitations
precisely match those of the classical solution (\ref{dark}), which
is why the holes, at arbitrary coupling, can be interpreted as
quantum dark solitons \cite{Kulish:1976ek}. The particle branch of
the spectrum interpolates between sound waves and bosonic
single-particle excitations and at $g\ll \sqrt{\mu }$ is described
by the Bogolyubov theory of a weakly interacting Bose gas
\cite{bogolyubov,AGD}.

The relationship between dark solitons and holes in the Fermi sea
essentially follows \cite{Kulish:1976ek,Smirnov:1982ke} from the
spectral properties of the Lax operator in the finite-density case
\cite{solitons}: the spectrum of the auxiliary linear problem has a
gap, which is the semiclassical counterpart of the Fermi sea in
quantum theory. The dark solitons (\ref{dark}) correspond to
normalizable eigenstates inside the gap and thus represent holes.
The spectral density has a characteristic square-root behavior at
the edges of the spectral gap. There are many instances where Bethe
Ansatz reduces to singular integral equations of the matrix model
type and its solutions exhibit similar square-root behavior. This
happens, for example, at large $N$
\cite{Fateev:1994dp,Fateev:1994ai}, in the semiclassical
approximation \cite{suther,Beisert:2003xu,Kazakov:2004qf} or in the
conformal limit
\cite{Mann:2005ab,Gromov:2006dh,Gromov:2006cq,Gromov:2007fn}. We
will demonstrate that Bethe equations also reduce to singular
integral equations when they describe weakly interacting Bose gas.
This limit of Bethe Ansatz (which we will call the Bogolyubov limit)
is ubiquitous in integrable systems, and does not necessarily
coincide with the classical approximation.

In particular, the large-$N$ limit of the $O(N)$ model, in which
quantum fluctuations are definitely important, falls into the
category described above. Building upon this observation we will
argue that quantum giant magnons should be identified with the holes
in the Fermi sea. We will first construct the large-$N$ counterpart
of the classical solution (\ref{HM}) in sec.~\ref{largeN} and then
compare it with the large-$N$ limit of Bethe Ansatz in
sec.~\ref{Bethe}. In sec.~\ref{XXZ} we study the limit of small
anisotropy in the XXZ spin chain which also turns out to be of the
Bogolyubov type.

\section{Giant magnons at large $N$}\label{largeN}

\subsection{$O(N)$ model at finite density}

In order to induce a finite density of one of the $O(N)$ charges
$Q_{ij}$ ($i,j=1\ldots N$) one can couple (\ref{on}) to a chemical
potential by shifting the Hamiltonian $H\rightarrow
H-\mu^{ij}Q_{ij}/2$. This is equivalent to gauging the $O(N)$
symmetry by a constant $A_0$ and amounts to replacing $\partial _0$
by a covariant derivative $ D^{ij} _0=
\partial _0\delta ^{ij}+\mu ^{ij}$ in the action. In the AdS/CFT
context the finite density of the $O(6)$ charge corresponds to an
infinite angular momentum uniformly distributed along the
string\footnote{See \cite{Roiban:2007ju} for a recent discussion of
the canonical vs. microcanonical description of charged states in
the AdS string theory}. The $N/2$ independent Cartan
charges\footnote{For simplicity we assume that $N$ is even. This
assumption is not essential in the large-$N$ limit.} are carried by
complex linear combinations
\begin{equation}\label{}
 z_I=\frac{n_{2I-1}+in_{2I}}{\sqrt{2}}\,,\qquad I=1\ldots N/2.
\end{equation}
Introducing a Lagrange multiplier $\sigma $ that enforces the
condition $z^{*I}z_I=1/2$, we can put (\ref{on}) into the
unconstrained form:
\begin{equation}\label{}
 S=\frac{N}{\lambda }\int_{}^{}d^2x\,
 \left[\sum_{I=1}^{N/2}\left(|D_\nu z_I|^2-\sigma |z_I|^2\right)
 +\frac{1}{2}\,\sigma \right],\qquad D_0z_I=\partial _0z_I-i\mu
 ^Iz_I.
\end{equation}
In principle all $\mu ^I$'s are independent variables, and one can
consider various combinations of the chemical potentials which give
different background charges. We will be interested in  the simplest
case when only one chemical potential $\mu \equiv \mu ^1$ is
non-zero and the rest $\mu ^I=0$.  For the field that carries the
background charge we will use a special notation\footnote{This is
the same $\phi $ as in (\ref{HM}).}:
\begin{equation}\label{nota}
 \phi \equiv \frac{1}{\sqrt{\lambda} }\,z_1.
\end{equation}

The large-$N$ limit of the $O(N)$ model can be solved by standard
methods \cite{Moshe:2003xn}. Integrating out $z_I$'s generates an
effective action for $\sigma $, which has a minimum at a non-zero
vev: $\left.\left\langle \sigma \right\rangle\right|_{\mu =0}=m^2 $.
The vev of $\sigma $ gives equal masses to all $z_I$ fields and is
determined by the gap equation:
\begin{equation}\label{gap}
 \frac{1}{\lambda }=\left\langle x\right|\frac{i}{-\partial ^2-m^2}
 \left| x\right\rangle.
\end{equation}

When the chemical potential is turned on it is convenient to leave
the charged field $\phi $ unintegrated:
\begin{eqnarray}\label{seff}
 S_{\rm eff}&=&N\int_{}^{}d^2x\,\left[
 |\partial _\nu \phi |^2+i\mu\left(
 \phi ^*\partial _0\phi -\phi\, \partial _0\phi ^*\right)
 +(\mu ^2-\sigma )|\phi |^2+\frac{1}{2\lambda }\,\sigma
 \right]\nonumber \\
 &&+\frac{i(N-2)}{2}\,\ln\det\left(-\partial ^2-\sigma \right).
\end{eqnarray}
At large $N$ the tree approximation for this effective action
becomes exact and one can expand around the minimum of the effective
potential. If $\mu
>m$, the setting is the same as in the Bogolyubov theory:
The zero mode of $\phi $ Bose condenses, with the physical ground
state at
\begin{equation}\label{}
 \left\langle \sigma \right\rangle=\mu ^2,\qquad \left\langle \phi
 \right\rangle^2=\frac{1}{4\pi }\,\ln\frac{\mu }{m}\,.
\end{equation}
These equations are obtained by minimizing the effective action
(\ref{seff}) in $\phi $ and $\sigma $ and taking into account the
dimensional transmutation formula (\ref{gap}). The value of the
action at the minimum determines the density of the free energy:
\begin{equation}\label{evac}
 \mathcal{E}=\frac{E_{\rm vac}}{{\rm Vol}}=-\frac{N\mu ^2}{8\pi }\left(2\ln\frac{\mu
 }{m}-1\right).
\end{equation}
The fluctuations of $\phi $ around the ground state (the Bogolyubov
branch of the spectrum) interpolate between phonons with
$\varepsilon=c_sp$, at $p\ll \mu \ln(\mu /m)$, and single-particle
excitations with $\varepsilon =p$, at $p\gg \max\{\mu \ln(\mu
/m),\mu \}$. The speed of sound can be found by integrating out
$\sigma $ in (\ref{seff}) and linearizing the resulting equations of
motion for $\phi $:
\begin{equation}\label{speedofs}
 c_s^2=\frac{\ln\frac{\mu }{m}}{\ln\frac{\mu }{m}+1}\,.
\end{equation}
Alternatively, the same result can be obtained from the
thermodynamic relation $c_s^2=\mu ^{-1}(\partial
\mathcal{E}/\partial \mu )/(\partial ^2\mathcal{E}/\partial \mu
^2)$. In addition to sound waves, the field $\phi $ describes a
massive mode separated from the ground state by the gap $M^2=8\mu
^2\ln(\mu /m)$. The neutral modes have a common mass equal to $\mu
$.

\subsection{Solitons}

We now turn to the soliton sector of the large-$N$ effective theory
(\ref{seff}). The effective action (\ref{seff}) and the ensuing
equations of motion are non-local\footnote{Here we used (\ref{gap})
to eliminate the cutoff dependence and to trade the bare coupling
for the physical mass. }:
\begin{eqnarray}\label{eqm}
 &&|\phi |^2=\frac{1}{2}\left\langle x\right|\left(
 \frac{i}{-\partial ^2-m^2}-\frac{i}{-\partial ^2-\sigma }\right)
 \left| x\right\rangle\nonumber \\
 &&-\partial ^2\phi +2i\mu \,\partial _0\phi +(\mu ^2-\sigma )\phi =0.
\end{eqnarray}
Similar equations arise in a variety of large-$N$ field theories and
in spite of their non-locality are solvable in some cases
\cite{Dashen:1975xh,Shei:1976mn,Feinberg:2003qz}, which presumably
reflects complete integrability of the underlying models. The $O(N)$
model is integrable as well and we will be able to construct the
exact giant magnon solution of (\ref{eqm}) by using a method
\cite{Feinberg:1994qq,Makeenko:1994pw,Feinberg:1994fq,Feinberg:2003qz}
based on the Gelfand-Diki\u{i} identities \cite{Gelfand:1975rn} for
the diagonal resolvent of the Sturm-Liouville operator:
\begin{equation}\label{GD}
 R\left[{\rm x};V({\rm x})\right]=\left\langle {\rm x}\right|\frac{1}{-\frac{d^2}
 {d{\rm x}^2}+V}\left| {\rm x}\right\rangle.
\end{equation}
With the help of the differential equation satisfied by the diagonal
resolvent \cite{Gelfand:1975rn} one can prove the following
remarkable identity:
\begin{equation}\label{gelfDik}
 R\left[{\rm x};\omega ^2-\frac{2\nu ^2}{\cosh^2\nu {\rm x}}\right]
 =\frac{1}{2\omega }+\frac{\nu ^2}{2\omega \left(\omega ^2-\nu ^2\right)\cosh^2\nu {\rm
 x}}\,.
\end{equation}

Since the giant magnon is a traveling dispersionless wave, it is
convenient to perform a Lorentz transformation to its rest
frame\footnote{In the presence of the background charge density the
Lorentz invariance is spontaneously broken, so this transformation
does not leave the equations of motion invariant.}:
\begin{equation}\label{}
 {\rm x}=\frac{x^1-vx^0}{\sqrt{1-v^2}}\,,\qquad {\rm
 t}=\frac{x^0-vx^1}{\sqrt{1-v^2}}\,.
\end{equation}
and to look for solutions independent of ${\rm t}$: $\sigma \equiv
\sigma ({\rm x}),~\phi \equiv \phi ({\rm x})$. After the Fourier
transform in ${\rm t}$, (\ref{eqm}) become:
\begin{eqnarray}\label{eqma}
 &&|\phi ({\rm x})|^2=\int_{-\infty }^{+\infty }\frac{d\omega }{4\pi
 }\,\,\left(
 R\left[{\rm x};\omega ^2+m^2\right]-R\left[{\rm x},\omega ^2+\sigma \right]
 \right)
 \nonumber \\
 &&\acute{\acute{\phi }}-\frac{2i\mu v}{\sqrt{1-v^2}}\,\acute{\phi
 }+(\mu ^2-\sigma )\phi =0.
\end{eqnarray}
The identity (\ref{gelfDik}) and the form of the classical
Hoffman-Maldacena solution (\ref{HM}) suggest the following ansatz:
\begin{eqnarray}\label{giantatlargen}
 \sigma &=&\mu ^2-\frac{2\nu ^2}{\cosh^2\nu {\rm x}}
 \nonumber \\
 \phi &=&\left({\frac{1}{4\pi }\,\ln\frac{\mu }{m}}\right)^{1/2}\,
 \frac{\mu v-i\nu \sqrt{1-v^2}\,\tanh\nu {\rm x}}{\sqrt{\nu ^2+(\mu ^2-\nu
 ^2)v^2}}\,.
\end{eqnarray}
\begin{figure}[t]
\centerline{\includegraphics[width=8cm]{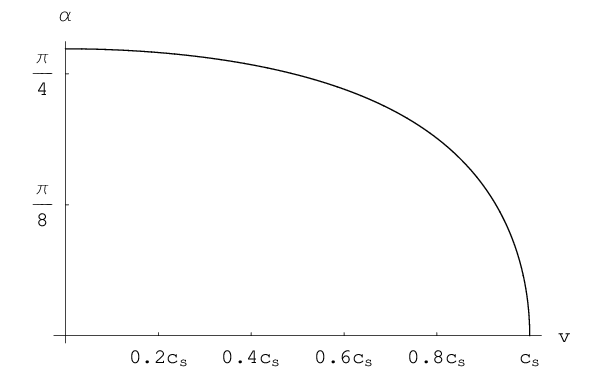}}
\caption{\label{aleff}\small The inverse size of the giant magnon as
a function of the velocity, at $\ln(\mu /m)=1$. The angle $\alpha $
is defined in (\ref{alffff})}
\end{figure}
It is straightforward to check that the ansatz goes through the
equations of motion (\ref{eqma}), provided that the ratio
\begin{equation}\label{alffff}
 \frac{\nu }{\mu }\equiv \sin\alpha
\end{equation}
satisfies
\begin{equation}\label{eqforalpha}
 \alpha \left(\tan\alpha +v^2\cot\alpha
 \right)=\left(1-v^2\right)\ln\frac{\mu }{m}\,.
\end{equation}

The last equation determines $\alpha $, and hence $\nu $, the
inverse size of the soliton, as a function of its velocity. The
function $\alpha (v)$ is plotted in fig.~\ref{aleff}. It reaches its
maximum at $v=0$ and then monotonously decreases with the increase
of $v$. In contradistinction to their classical counterparts, the
large-$N$ giant magnons cannot move faster than sound\footnote{The
classical approximation is accurate at asymptotically high densities
when $\ln(\mu /m)$ is large and according to (\ref{speedofs}) the
speed of sound approaches one. The limiting velocity thus is not
visible in the classical approximation.}: when $v$ approaches $c_s$,
defined in (\ref{speedofs}), $\alpha $ goes to zero. The soliton
becomes larger and larger and completely dissociates when $v=c_s$.

In the weak-coupling limit, $\mu \gg m$, the large-$N$ solution
(\ref{giantatlargen})-(\ref{eqforalpha}) goes over to the classical
giant magnon (\ref{HM}), because then $\alpha \approx \pi /2$
(unless the velocity is very close to the speed of sound) and
consequently $\nu \approx \mu $. An overall logarithmic factor in
(\ref{giantatlargen}) arises because of the different normalization
of the field $\phi $, eq.~(\ref{nota}). The bare coupling $\lambda $
there gets replaced by the running coupling at the scale $\mu $:
$\lambda \rightarrow 2\pi /\ln(\mu /m)$.

To calculate the energy and the momentum of the giant magnon, we
first compute the effective Lagrangian:
\begin{equation}\label{}
 T\mathcal{L}={S_{\rm eff}[\phi ,\sigma ]-S_{\rm vac}}.
\end{equation}
The vacuum term subtracts the bulk energy and the momentum of the
background state without the soliton. It is a bit surprising that
the vacuum carries not only the bulk energy but also a finite amount
of momentum. The non-zero momentum arises because the soliton
belongs to a sector with twisted boundary conditions. The phase of
$\phi $ experiences a non-zero increment on the soliton solution
(\ref{giantatlargen}):
\begin{equation}\label{deltaphi}
 \Delta \varphi =-2\arctan\frac{\nu \sqrt{1-v^2}}{\mu v}\,.
\end{equation}
 Consequently, the ground state must be  position-dependent,
 in order to satisfy the boundary conditions:
\begin{equation}\label{phivac}
 \phi_{\rm vac} =\left(\frac{1}{4\pi }\,\ln\frac{\mu }{m}\right)^{1/2}
 \,{\rm e}\,^{i\Delta \varphi\sqrt{1-v^2}\,{\rm x}/L},
\end{equation}
where $L$ is the size of the system. At $L\rightarrow \infty $ the
phase changes so slowly that it does not contribute to the energy,
but it still contributes a finite amount to the momentum (the
momentum density due to the phase rotation is $O(1/L)$, while the
energy density is $O(1/L^2)$). The details of the calculation can be
found in appendix~\ref{act}. The result is
\begin{equation}\label{Lagrange}
 \frac{\pi }{N\mu }\,\mathcal{L}(v)=\ln\frac{\mu }{m}\, v\arctan\frac{\sqrt{1-v^2}\,\sin\alpha
 }{v}-\sqrt{1-v^2}\left[
 \left(\ln\frac{\mu }{m}-1\right)\sin\alpha +\alpha \cos\alpha
 \right].
\end{equation}
The energy and momentum of the giant magnon can now be found from
\begin{equation}\label{}
 p=\frac{d\mathcal{L}}{dv}\,,\qquad \varepsilon =pv-\mathcal{L}.
\end{equation}
The calculation is simplified by the fact that $\partial
\mathcal{L}/\partial \alpha =0$  as long as (\ref{eqforalpha}) is
satisfied, so that $d\mathcal{L}/dv=\partial \mathcal{L}/\partial
v$:
\begin{eqnarray}\label{p}
 \frac{\pi }{N\mu }\,p&=&\ln\frac{\mu }{m}\,\arctan\frac{\sqrt{1-v^2}\,\sin\alpha
 }{v}-\frac{v\sin\alpha }{\sqrt{1-v^2}} \\ \label{e}
 \frac{\pi }{N\mu }\,\varepsilon &=&\frac{\alpha \sec\alpha -\sin\alpha
 }{\sqrt{1-v^2}}\,.
\end{eqnarray}
These two equations, together with (\ref{eqforalpha}), determine the
dispersion relation $\varepsilon =\varepsilon (p)$ of the giant
magnon in an implicit form.

Contrary to naive expectations, the momentum of the giant magnon
decreases with increasing velocity.  Since $\alpha =0$ at $v=c_s$,
the magnon moving at the speed of sound has zero momentum and zero
energy. As the velocity approaches zero, the momentum reaches its
maximal value
\begin{equation}\label{}
 p_F=\frac{N\mu  }{2}\,\ln\frac{\mu }{m}\,,
\end{equation}
which we will call the Fermi momentum for the reasons that will
become clear in the next section. The energy is a periodic function
of the momentum with the period $2p_F$:
\begin{equation}\label{}
 \varepsilon (p+2p_F)=\varepsilon (p),
\end{equation}
because of the ambiguity in choosing the branch of the arctangent in
(\ref{p}). The momentum is thus naturally confined within a single
"Brillouin zone" $-p_F<p<p_F$.

\begin{figure}[t]
\centerline{\includegraphics[width=10cm]{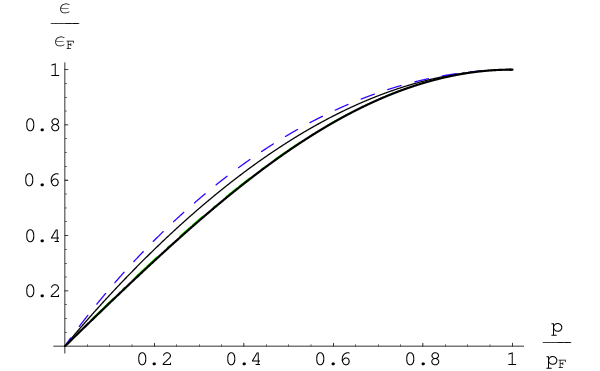}}
\caption{\label{kart1}\small The normalized dispersion relation of
the giant magnon: the thick solid curve is the $\sin$ law,
eq.~(\ref{appro}); the exact dispersion relation  is practically
indistinguishable from it already at $\ln(\mu /m)=10$ (dot-dashed
green line). The thin solid line corresponds to $\ln(\mu /m)=1$ and
the dashed blue line to $\ln(\mu /m)=0.01$.}
\end{figure}

The Fermi momentum and the Fermi energy grow logarithmically with
$\mu $. At very large $\mu $:
\begin{equation}\label{}
 p_F=\frac{\pi N\mu }{\lambda (\mu )}\,,\qquad
 \varepsilon _F\approx \frac{2 N\mu }{\lambda (\mu )}\qquad (\mu \rightarrow \infty
 ),
\end{equation}
where $\lambda (\mu )=2\pi /\ln(\mu /m)$ is the running coupling.
The limit of large chemical potential is the weak-coupling
perturbative limit. The second term on the right-hand-side of
(\ref{p}) can then be neglected. Also $\alpha$ approaches $\pi /2$,
and the dispersion relation becomes
\begin{equation}\label{appro}
 \varepsilon (p)\approx \varepsilon _F\sin\frac{\pi p}{2p_F}\qquad (\mu \rightarrow \infty ).
\end{equation}
The periodicity in momentum is manifest here. In the classical
approximation it has a nice geometric interpretation
\cite{Hofman:2006xt}: The momentum of the classical giant magnon is
the angle subtended by the ends of the string on the sphere. It is
interesting that the momentum gets non-geometric quantum corrections
already in the large-$N$ approximation. The geometric $\arctan$ term
in (\ref{p})), of order $1/\lambda (\mu )$, is shifted by a quantum
term, of order one, which has no apparent geometric meaning.
Numerically, (\ref{appro}) is a good approximation in the whole
range of parameters, as can be seen from fig.~\ref{kart1}.

\section{Bethe Ansatz}\label{Bethe}

The exact quantum spectrum of the $O(N)$ model consists of $N$
massive particles in the vector representation of $O(N)$, whose
S-matrix is known exactly at any $N$ \cite{Zamolodchikov:1978xm}.
The ground state at finite density is the Fermi sea of fundamental
particles that occupy a finite rapidity interval. The distribution
of particles in the ground state at non-zero chemical potential and
zero temperature is given by the solution of the following integral
equation \cite{Hasenfratz:1990zz,Hasenfratz:1990ab}:
\begin{equation}\label{tba}
 \varepsilon (\theta )-\int_{-B}^{B}d\xi \,K(\theta -\xi )\varepsilon
 (\xi )=m\cosh\theta -\mu.
\end{equation}
The kernel $K(\theta )$ is the derivative of the scattering phase
shift that can be extracted from the exact S-matrix by taking the
matrix element responsible for scattering of particles that carry
the background charge (at large $N$ these are the quanta of the
field $\phi $ in (\ref{seff})). The kernel is a rather involved
function of the relative rapidity \cite{Hasenfratz:1990ab}:
\begin{eqnarray}\label{kern}
 K(\theta )&=&\frac{1}{4\pi ^2}
 \left[
 \psi \left(\frac{i\theta }{2\pi}\right)
 -\psi \left(\frac{1}{N-2}+\frac{i\theta }{2\pi}\right)
 +\psi \left(\frac{1}{2}+\frac{1}{N-2}+\frac{i\theta }{2\pi}\right)
 -\psi \left(\frac{1}{2}+\frac{i\theta }{2\pi}\right)
 \right.\nonumber \\
 &&\left.+\psi \left(-\frac{i\theta }{2\pi}\right)
 -\psi \left(\frac{1}{N-2}-\frac{i\theta }{2\pi}\right)
 +\psi \left(\frac{1}{2}+\frac{1}{N-2}-\frac{i\theta }{2\pi}\right)
 -\psi \left(\frac{1}{2}-\frac{i\theta }{2\pi}\right)
 \right].
\end{eqnarray}
The equation (\ref{tba}) describes the filling of the Fermi sea in
the thermodynamic limit. The rapidity interval $(-B,B)$ is occupied,
while the outside of the interval is empty. The equation describes
not only the ground state, but also the spectrum of
excitations\footnote{The equation (\ref{tba}) does not take into
account the spin degrees of freedom. The nested Bethe equations,
which describe spins of the particles, are analyzed for the $O(N)$
model at finite density in
\cite{Gromov:2006dh,Gromov:2006cq,Gromov:2007fn}. It is interesting
that the spin excitations have much in common with giant magnons
\cite{Gromov:2006cq}.}. The ground state energy is given by
\begin{equation}\label{tbafree}
 \mathcal{E}=\frac{m}{2\pi }\int_{-B}^{B}d\theta \,\varepsilon (\theta )
 \cosh\theta .
\end{equation}
The function $\varepsilon (\theta )$, often called pseudo-energy, is
the energy of a particle (for $|\theta |>B$) or a hole (for $|\theta
|<B$) with rapidity $\theta $. Consequently,  $\varepsilon
(\theta)\lessgtr 0$ at $|\theta |\lessgtr B$. The condition that
$\varepsilon (\pm B)=0$ unambiguously determines the Fermi rapidity
$B$.

We will be mostly interested in the hole excitations. To find their
dispersion relation $\varepsilon =\varepsilon (p)$ one has to solve
an additional equation \cite{inverse}:
\begin{equation}\label{ptba}
 \acute{p}(\theta )-\int_{-B}^{B}d\xi \,K(\theta -\xi )\acute{p}
 (\xi )=-m\cosh\theta,
\end{equation}
which determines (the derivative of) the momentum. We are going to
show that the hole excitations  are equivalent at $N\rightarrow
\infty $ to the solitons constructed in previous section. It is
instructive to consider first a much simpler case of the NLS model,
where the holes in the Fermi sea can be shown to describe quantum
dark solitons.

\subsection{Non-linear Schr\"odinger model}

The integral equations for the NLS model (\ref{NLS}) are
\cite{Lieb:1963rt,Yang:1968rm,inverse}
\begin{eqnarray}\label{enls}
 \varepsilon (v )-\frac{g}{\pi }\int_{-B}^{B}
 \frac{du \,\varepsilon (u )}{(v -u )^2+g^2}&=&v
 ^2-\mu
 \\
 \label{pnls}
 \acute{p}(v )-\frac{g}{\pi }\int_{-B}^{B}
 \frac{du \,\acute{p}(u )}{(v -u )^2+g^2}&=&-1.
\end{eqnarray}
These equations are much simpler than the equations for the ground
state of the $O(N)$ model, yet they are not solvable analytically.
In \cite{Lieb:1963rt} they were analyzed numerically. Not
surprisingly the equations considerably simplify in the Bogolyubov
limit $g\rightarrow 0$, such that they admit an analytic solution.
At first sight, the kernel simply disappears at $g\rightarrow 0$,
because the scattering phase is then very small. Neglecting the
kernel, however, would lead to totally misleading results, because
the scattering phase is small only for $|v-u|\gg g$. If $|u-v|\sim
g$ the kernel on the contrary is very large: $K\sim 1/g$. In fact,
$K(v)$ approximates the delta-function at small $g$. But replacing
$K(v)$ by $\delta (v)$ would again be wrong\footnote{Such an
approximation is correct at finite temperature or, more precisely at
$T\gg \mu ^{3/2}/g$, and leads to the standard Bose distribution
\cite{Yang:1968rm}.}, since then the left-hand side of (\ref{enls})
completely disappears. The correct procedure consists in keeping the
next-to-leading $O(g)$ term\footnote{This is only important for
holes. For particles the delta-function is concentrated outside of
the region of integration.}:
\begin{equation}\label{exapns}
 \frac{g}{v ^2+g^2}\approx \pi \,\delta (v )
 +\wp\frac{g}{v ^2}\qquad (g\rightarrow 0).
\end{equation}
The equation for the pseudo-energy of holes then becomes a singular
integral equation:
\begin{equation}\label{}
 \frac{g}{\pi }
 -\!\!\!\!\!\!\!\int_{-B}^{B}\frac{du \,\varepsilon (u )}{(v -u )^2}
 =\mu -v ^2\qquad (|v|<B).
\end{equation}
Integrating once we get:
\begin{equation}\label{efinNLS}
 -\frac{g}{\pi }-\!\!\!\!\!\!\!\int_{-B}^{B}
 \frac{du \,\varepsilon (u )}{v -u}
 =\mu v -\frac{1}{3}\,v ^3,
\end{equation}
and, similarly,
\begin{equation}\label{pfinNLS}
 -\frac{g}{\pi }-\!\!\!\!\!\!\!\int_{-B}^{B}
 \frac{du \,\acute{p}(u )}{v -u}
 =v .
\end{equation}
These equations determine the energy and momentum of holes, which thus scale as $1/g$. 

The energy and momentum
of particles can be also calculated, directly from (\ref{enls}), (\ref{pnls}):
\begin{eqnarray}\label{}
 \varepsilon (v)&=& v^2-\mu+\frac{g}{\pi }\int_{-B}^{B}
 \frac{du \,\varepsilon (u )}{(v -u )^2}\qquad (|v|>B)
 \nonumber \\
 \acute{p}(v)&=&1-\frac{g}{\pi }\int_{-B}^{B}
 \frac{du \,\acute{p}(u )}{(v -u )^2}\qquad (|v|>B),
\end{eqnarray}
where $\varepsilon (u)$, $\acute{p}(u)$ in the integrands are the energy and momentum of holes. Since they scale as $1/g$, the energy and momentum of particles are of order one in the weak-coupling limit.

The equation (\ref{efinNLS}) is easily solvable. It also admits an
interesting interpretation in terms of random matrix theory, where
such an equation arises as an equilibrium condition for an
eigenvalue distribution \cite{Brezin:1977sv}, which can be pictured
as a macroscopically large number of particles in an external
potential $V_{\rm ext}=\mu v^2/2-v^4/6$ subject to pairwise
logarithmic repulsion. The equation itself does not determine the
Fermi velocity $B$. In matrix models the normalization of the
density $-\varepsilon (v)$ uniquely determines $B$
\cite{Brezin:1977sv}, but here the total number of particles is not
fixed and in order to find the Fermi velocity we need to minimize
the free energy:
\begin{equation}\label{freee}
 \mathcal{E}=\frac{1}{2\pi }\int_{-B}^{B}dv\,\varepsilon (v).
\end{equation}
In the matrix-model language, $\mathcal{E}$ is the total number of
particles (up to a sign since $\varepsilon (v)$ is negative inside
the Fermi interval and thus $\mathcal{E}<0$). Therefore, we need to
increase the number of particles as much as possible in order to
minimize $\mathcal{E}$. This cannot be done indefinitely because the
potential $V_{\rm ext}$ has the shape of an upside-down double well.
The repulsion between the particles counteracts the attraction
towards the bottom of the potential and tends to spread the
particle's distribution. Eventually, if the number of particles is
sufficiently large, the repulsion wins and the particles start to
spill out of the potential well. Mathematically this means that for
sufficiently large $B$, eq.~(\ref{efinNLS}) has no solutions with
$\varepsilon (v)<0$ within the Fermi interval $(-B,B)$. The free
energy is minimized by the critical solution, when the particles are
just starting to spill out of the potential well. The critical point
is characterized by the change in the edge behavior of the
particle's density. Normally $\varepsilon (v)\sim (B-v)^{1/2}$, but
at the critical point \cite{Brezin:1977sv}
\begin{equation}\label{edge}
 \varepsilon (v)\sim (B-v)^{3/2}.
\end{equation}
The criticality gives an extra condition that determines $B$.
Imposing this condition on the solution of (\ref{efinNLS}), we find\footnote{In \cite{Tyupkin:1975pk} this solution was obtained by expanding in the basis of Chebyshev polynomials, which diagonalize the Hilbert transform.}:
\begin{eqnarray}\label{}
 \varepsilon (v)&=&\frac{1}{3g}\,\left(2\mu -v^2\right)^{3/2}
 \\
 p(v)&=&\frac{\mu }{g}\,\arctan\frac{\sqrt{2\mu
 -v^2}}{v}-\frac{v}{2g}\,\sqrt{2\mu -v^2}
 \\
 \mathcal{E}&=&-\frac{\mu ^2}{4g}\,.
\end{eqnarray}
These are, respectively, the energy of the dark soliton (\ref{dark})
\cite{Kulish:1976ek}, its momentum \cite{Kulish:1976ek}, and the
energy density of the ground state at $\left\langle \phi
\right\rangle=\sqrt{\mu /2g}$.

\subsection{$O(N)$ sigma model}

The large-$N$ limit of the Bethe Ansatz in the $O(N)$ model is very
similar to the weak coupling limit for NLS. The large-$N$ expansion
of the kernel (\ref{kern}) starts with the delta-function. Keeping
the next-to-leading term, we get:
\begin{equation}\label{}
 K(\theta )\approx \delta (\theta )+\frac{1}{N}\,\wp
 \left(\frac{\cosh\theta }{\sinh^2\theta }+\frac{1}{\theta
 ^2}\right).
\end{equation}
Repeating the same steps as in the NLS case, we arrive at the
singular integral equations for the pseudo-energy of holes:
\begin{equation}\label{eN}
 -\frac{1}{N}-\!\!\!\!\!\!\!\int_{-B}^{B}d\xi \,\varepsilon (\xi  )
 \left(\frac{1}{\theta -\xi }+\frac{1}{\sinh(\theta -\xi )}\right)
 =\mu \theta -m\sinh\theta ,
\end{equation}
and for their momentum:
\begin{equation}\label{pN}
 -\frac{1}{N}-\!\!\!\!\!\!\!\int_{-B}^{B}d\xi \,\acute{p}(\xi )
 \left(\frac{1}{\theta -\xi }+\frac{1}{\sinh(\theta -\xi )}\right)
 =m\sinh\theta .
\end{equation}

The singular integral equations with a combination of  rational and
hyperbolic kernels are not solvable by standard techniques, but as
we will argue, the solution is implicitly given by the dispersion
relation of the large-$N$ giant magnon, eqs.~(\ref{p}), (\ref{e}),
(\ref{eqforalpha}). It is straightforward to check this
perturbatively in $B$ and $\theta $, which is effectively an
expansion in $\ln(\mu /m)$. The tricky part is to find the
relationship between the rapidity $\theta $, that enters the Bethe
equations, and the velocity of the giant magnon, or the parameter
$\alpha $ defined in (\ref{eqforalpha}). To the first few orders in
$B$,
\begin{equation}\label{}
 \alpha = \left(
 \frac{1}{2}
 -\frac{B^2}{24}
 +\frac{B^4}{144}+\frac{B^2\theta ^2}{360}
 +\ldots
 \right)\sqrt{B^2-\theta ^2}\,.
\end{equation}
Then
\begin{eqnarray}\label{etban}
 \varepsilon (\theta )&=&-\frac{N\mu }{12\pi }\left(
 1
 -\frac{7B^2}{40}+\frac{\theta ^2}{20}
 +\frac{527B^4}{13440}-\frac{3B^2\theta ^2}{280}+\frac{\theta ^4}{840}
 +\ldots
 \right)\left(B^2-\theta ^2\right)^{3/2}
 \nonumber \\
 p(\theta )&=&\frac{N\mu }{4\pi }\left[\left(
 B^2
 -\frac{B^4}{12}
 +\frac{11B^6}{720}
 +\ldots
 \right)\arccos\frac{\theta }{B}
 \right.
 \nonumber \\
 &&\left.
 -\left(
 \theta
 -\frac{B^2\theta }{6}+\frac{\theta ^3}{12}
 +\frac{11B^4\theta }{320}-\frac{7B^2\theta^3 }{320}+\frac{\theta ^5}{360}
 +\ldots
 \right)\sqrt{B^2-\theta ^2}
 \right]
\end{eqnarray}
indeed solve (\ref{eN}) and (\ref{pN}) provided that
\begin{equation}\label{muover}
 \frac{\mu }{m}=1+\frac{B^2}{4}+\frac{B^4}{96}+\ldots \,.
\end{equation}

This perturbative solution can be pushed to any reasonable order
using {\tt Mathematica} and passes a number of consistency checks:
The free energy computed from (\ref{tbafree}):
\begin{equation}\label{}
 \mathcal{E}=\frac{N\mu ^2}{8\pi }\left(1
 -\frac{B^2}{2}+\frac{B^4}{24}-\frac{11B^6}{1440}+\ldots \right)
\end{equation}
agrees with (\ref{evac}) upon identification (\ref{muover}).

Since $\varepsilon (\theta )\sim (B-\theta )^{3/2}$, we can
differentiate (\ref{eN}) in $m $ without the risk of producing a
singularity at the edge the Fermi interval. This gives the
relationship:
\begin{equation}\label{kin}
 \acute{p}=-m\,\frac{\partial \varepsilon }{\partial m }\,,
\end{equation}
which is also compatible with the solution (\ref{etban}).

The last equation can be used to calculate the exact Fermi rapidity.
Near the Fermi point $\theta =B$, the pseudo-energy has the form
$\varepsilon (\theta )=-P(m,\theta )(B-\theta )^{3/2}$, where
$P(m,\theta )$ is analytic at $\theta =B$. Differentiating in $m$,
we find from (\ref{kin}):
$$
 \acute{p}(\theta )=\frac{3}{2}\,mP(m,B)\,\frac{\partial B}{\partial
 m}\,\left(B-\theta \right)^{1/2}+O\left((B-\theta )^{3/2}\right),
$$
or
$$
 {p}(\theta )=-mP(m,B)\,\frac{\partial B}{\partial
 m}\,\left(B-\theta \right)^{3/2}+O\left((B-\theta )^{5/2}\right).
$$
The ratio $\varepsilon /p$ at the Fermi point coincides with the
speed of sound:
\begin{equation}\label{}
 c_s=-\lim_{\theta \rightarrow B}\frac{\varepsilon (\theta )}{p(\theta
 )}=-\frac{1}{m\,\frac{\partial B}{\partial m}}\,.
\end{equation}
Equating this to (\ref{speedofs}) gives a differential equation that
determines $B$:
\begin{equation}\label{exactB}
 B=\sqrt{\ln\frac{\mu }{m}\left(\ln\frac{\mu }{m}+1\right)}
 +\mathop{\rm arcsinh}\sqrt{\ln\frac{\mu }{m}}\,.
\end{equation}
Inverting this equation and expanding in $B$, we find
(\ref{muover}).

\section{Nearly isotropic XXZ spin chain}\label{XXZ}

In this section we consider the XXZ spin chain in the magnetic
field:
\begin{equation}\label{}
 H_{XXZ}=-\sum_{l=1}^{L}\left(\sigma _l^x\sigma _{l+1}^x
 +\sigma _l^y\sigma _{l+1}^y+\cos 2\eta \,\sigma _l^z\sigma _{l+1}^z+h\sigma
 _l^z\right).
\end{equation}
The ground state is described by the integral equations
\cite{inverse}
\begin{eqnarray}\label{}
 &&\varepsilon (\lambda )-\int_{-B}^{B}
 d\nu  \,K(\lambda -\nu )\varepsilon (\nu )=\varepsilon _0(\lambda )
 \\
 &&\acute{p}(\lambda )-\int_{-B}^{B}
 d\nu  \,K(\lambda -\nu )\acute{p}(\nu )
 =-\acute{p}_0(\lambda )
 \\ \label{freexxz}
 &&
 \mathcal{E}=\frac{1}{2\pi }\int_{-B}^{B}d\lambda \,\acute{p}_0(\lambda )
 \varepsilon(\lambda ),
\end{eqnarray}
where
\begin{eqnarray}\label{xxzkern}
 K(\lambda )&=&\frac{\sin 4\eta }
 {2\pi \sinh(\lambda +2i\eta )\sinh(\lambda -2i\eta )}
 \\
 \varepsilon _0(\lambda )&=&
 2h-\frac{2\sin^22\eta }{\cosh(\lambda +i\eta )\cosh(\lambda -i\eta
 )}
 \\
 \acute{p}_0(\lambda )&=&\frac{\sin 2\eta }{\cosh(\lambda +i\eta )\cosh(\lambda -i\eta
 )}\,.
\end{eqnarray}

The limit of small anisotropy, $\eta \rightarrow 0$, and small
magnetic field $h\sim \eta ^2$ can be interpreted as the Bogolyubov
limit. Indeed the small-$\eta$ expansion of the kernel
(\ref{xxzkern}) starts with the delta function:
\begin{equation}\label{}
 K(\lambda )\approx \delta (\lambda )+\wp\frac{2\eta }{\pi \sinh^2\lambda }\qquad
 (\eta \rightarrow 0),
\end{equation}
and Bethe Ansatz reduces to singular integral equations:
\begin{figure}[t]
\centerline{\includegraphics[width=8cm]{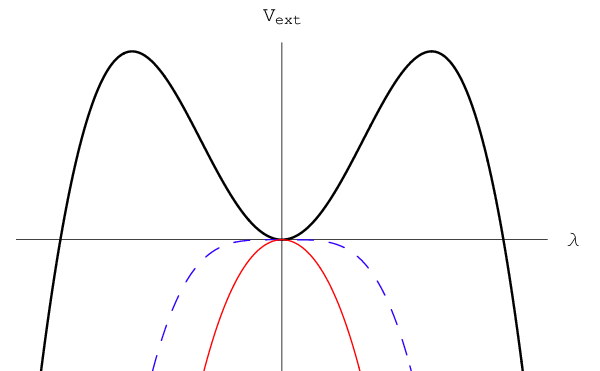}}
\caption{\label{epotxxz}\small The effective potential for the
integral equation (\ref{enrg}): at $h=0.8h_c$ (thick solid black
curve); at $h=h_c$ (dashed blue curve); and at $h=1.2h_c$ (thin
solid red curve).}
\end{figure}
\begin{eqnarray}\label{enrg}
 --\!\!\!\!\!\!\!\int_{-B}^{B}\frac{d\nu}{\pi }\, \,\varepsilon (\nu )\coth(\lambda -\nu )
 &=&4\eta \tanh\lambda -\frac{h}{\eta }\,\lambda .
 \\
 -\!\!\!\!\!\!\!\int_{-B}^{B}\frac{d\nu}{\pi }\, \,\acute{p}(\nu )\coth(\lambda -\nu )
 &=&-\tanh\lambda.
\end{eqnarray}
The effective "matrix-model" potential in (\ref{enrg}), $V_{\rm
ext}=4\eta \ln\cosh\lambda -h\lambda ^2/2\eta $, has a stable
minimum only if $h<4\eta ^2$. At the critical magnetic field
$h_c=4\eta ^2$ the minimum disappears (fig.~\ref{epotxxz}), the
Fermi interval shrinks to a point,
 and for
$h>h_c$ the equation(\ref{enrg}) has no solutions with negative
pseudo-energy.  The ground state at a supercritical magnetic field
is the completely empty ferromagnetic vacuum.

The $\coth$ kernel in (\ref{enrg}) can be explicitly inverted. After
straightforward albeit lengthy calculations we find the solution to
(\ref{enrg}) at criticality:
\begin{equation}\label{}
 \varepsilon (\lambda )=-\frac{1}{\eta \cosh\lambda }\,\sqrt{16\eta ^4-h^2\cosh^2\lambda }
 +\frac{h}{2\eta }\,\arccos\left(\frac{h^2}{8\eta ^4}\,\cosh^2\lambda
 -1\right),
\end{equation}
where the Fermi point is given by
$$
 \cosh B=\frac{4\eta ^2}{h}\,.
$$
One can verify that the pseudo-energy is negative everywhere in the
interval $(-B,B)$ and behaves as $|B\pm \lambda|^{3/2}$ at the
edges.

From (\ref{freexxz}) we get for the free energy density:
\begin{equation}\label{}
 \mathcal{E}=-\frac{(4\eta ^2-h)^2}{8\eta
 ^2}=-\frac{(h_c-h)^2}{2h_c}\,.
\end{equation}
The momentum can be computed by noticing that
\begin{equation}\label{}
 \acute{p}=\frac{\partial\left(\varepsilon \eta \right) }{\partial
 h_c}\,,
\end{equation}
which gives:
\begin{equation}\label{}
 p(\lambda )=\arccos\frac{h_c\tanh\lambda }{\sqrt{h_c^2-h^2}}
 -\frac{h}{h_c}\,\arccos\frac{h\sinh\lambda }{\sqrt{h_c^2-h^2}}\,.
\end{equation}
\begin{figure}[t]
\centerline{\includegraphics[width=8cm]{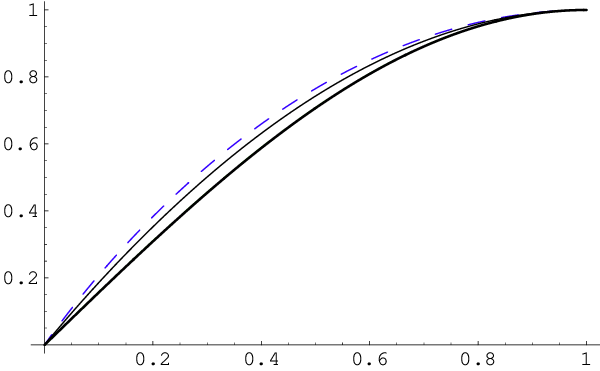}}
\caption{\label{xxzdis}\small The dispersion relation for dark
soliton in the XXZ spin chain for various values of the magnetic
field: $h_c/h=100$ (green dot-dashed line); $h_c/h=2$ (thin solid
line) and $h_c/h=1.01$ (dashed blue line). It is accurately fitted
by a simple dispersion law (\ref{appro}) shown in thin black line.}
\end{figure}
The velocity of sound is
\begin{equation}\label{}
 c_s=-\lim_{\lambda \rightarrow B}\frac{\varepsilon (\lambda )}{p(\lambda )}
 =\frac{\sqrt{h_c^2-h^2}}{\eta }\,.
\end{equation}
The dispersion relation is shown in fig.~\ref{xxzdis} and is well
approximated by (\ref{appro}), especially for small values of the
magnetic field.

The energy of a hole (which can presumably be identified with some
sort of a soliton) is periodic in momentum with the period $2p_F$,
where the Fermi momentum is given by
\begin{equation}\label{}
 p_F=\frac{\pi }{2}\,\,\frac{h_c-h}{h_c}\,.
\end{equation}
For very small magnetic fields the period is just the inverse of the
lattice spacing ($=1$). This is not surprising, since at zero
magnetic field the hole excitations are magnons of the XXZ spin
chain. Periodicity of their momentum is a consequence of the
underlying lattice structure. However, the effective lattice
spacing, $a_{\rm eff}=h_c/(h_c-h)$,  grows with the magnetic field
and becomes infinite at the critical point. The periodicity in
momentum should have some other origin near the critical point, not
related to the lattice structure of the spin chain.

\section{Conclusions}\label{conlusions}

The giant magnons in the $O(N)$ sigma-model, as well as other dark
soliton in integrable theories, can be identified with the holes in
the Fermi sea. The mysterious periodicity of their momentum has a
rather mundane explanation from this point of view -- the period is
just the Fermi momentum doubled. It is not clear what implications
can have such an interpretation for the AdS/CFT correspondence.
Unlike the string sigma-model, the $O(N)$ model is not conformal, it
is a massive field theory with non-zero beta-function and
dimensional transmutation. In addition, the string sigma-model is
coupled to 2d gravity and one should fix the diffeomorphism gauge
and solve or impose the Virasoro constraints\footnote{See
\cite{Gromov:2007fn} for a discussion of the Virasoro constraints
from the Bethe-Ansatz point of view in the conformal limit of the
$O(N)$ model.}. This eliminates longitudinal degrees of freedom,
which in the $O(N)$ model correspond to the Bogolyubov sound waves.
The giant magnons, however are transverse since they satisfy the
Virasoro constraints \cite{Hofman:2006xt} at least classically.

In string theory, the finite charge density arises when a physical
gauge condition of light-cone type is imposed. The zero-density
state and the spectrum of excitations around it presumably
correspond to the covariant, conformal-gauge description of the
sigma-model on $AdS_5\times S^5$, which at the moment is not
developed to the degree that one could formulate Bethe Ansatz in
the bare vacuum.

\subsection*{Acknowledgments}
I would like to thank G.~Arutyunov, V.~Fateev, N.~Gromov, V.~Kazakov,
C.~Kristjansen, Yu.~Makeenko, J.~Minahan, O.~Ohlsson~Sax,
F.~Smirnov, P.~Vieira and A.~Zamolodchikov for interesting
discussions and the Isaac Newton Institute for Mathematical Sciences
for kind hospitality during the programme "Strong Fields,
Integrability and Strings". This work was supported by the Knut and
Alice Wallenberg Foundation through the Royal Swedish Academy of
Sciences Research Fellowship, in part by the Swedish Research
Council under contract 621-2007-4177, and  in part by the Isaac
Newton Institute for Mathematical Sciences.


\appendix

\section{Action of the giant magnon}\label{act}

Here we compute the action of the large-$N$ giant magnon. All the
"classical" terms in the effective action (\ref{seff}), those that
depend on $\phi $, do not contribute since upon integration by parts
they yield the equation of motion for $\phi $, of which the giant
magnon is a solution. The easiest way to compute the "quantum" part
of the action is to differentiate it in $\nu $:
\begin{eqnarray}\label{}
 \frac{\partial S}{\partial \nu }&=&\frac{N}{2}
 \int_{}^{}d^2x\,\,\frac{\partial \sigma }{\partial \nu }
 \left(
 \frac{1}{\lambda }-\left\langle x\right|\frac{i}{-\partial ^2-\sigma }\left| x\right\rangle
 \right)\nonumber \\
 &=&
 TN\sqrt{1-v^2}\int_{}^{}d{\rm x}\,\,\frac{\partial \sigma }{\partial \nu
 }\,\int_{-\infty }^{+\infty }\frac{d\omega }{4\pi }\,\,\left(
 R\left[{\rm x};\omega ^2+m^2\right]-R\left[{\rm x};\omega ^2+\sigma
 \right]\right)\nonumber \\
 &=&\frac{1}{\pi }\left(
 \alpha \tan\alpha -\ln\frac{\mu }{m}
 \right), \nonumber
\end{eqnarray}
where in the last line we used (\ref{gelfDik}) and the explicit form
of the solution (\ref{giantatlargen}). Requiring that $S(\nu =0)=0$
effectively subtracts the background energy (but not the background
momentum!), and yields:
$$
 S=-\frac{1}{\pi }\,TN\mu \sqrt{1-v^2}\left[
 \left(\ln\frac{\mu }{m}-1\right)\sin\alpha+\alpha \cos\alpha
 \right].
$$
The background action due to the phase rotation in (\ref{phivac}) is
$$
 S_{\rm vac}=\frac{1}{2\pi }\,TN\mu v\,\Delta \varphi \,\ln\frac{\mu
 }{m}\,.
$$
Subtracting $S_{\rm vac}$ from $S$, and using (\ref{deltaphi}), we
get (\ref{Lagrange}).

\bibliographystyle{nb}

\end{document}